\documentclass[12pt,a4paper]{article}
\usepackage{amsmath}
\usepackage{amssymb}
\usepackage{amsfonts}
\usepackage{wasysym}
\usepackage{cite}
\title{Caducity of  idea about wave function collapse as well new views on Schrodinger's cat and quantum measurements}
\author{Spiridon Dumitru 
\footnote{(Retired)Department of Physics, "`Transilvania"' University, B-dul Eroilor 29, 500036 Brasov,  Romania, Phone: +40 746 058 152,  E-mail: s.dumitru42@yahoo.com}}
\date{\today}
\begin{document}
\maketitle
\begin{abstract}
Investigated idea was actuated by the old opinion that a measurement of a quantum observable  should be regarded a as a single deterministic sampling. But, according to the last decades studies, such observables are  veritable random variables and their measurements must imply significant sets of statistical samplings. So one finds the indubitable caducity of the approached idea.  Contiguously the respective finding allows to put into a new light the  controversial questions like the Schrodinger's cat thought experiment or description of quantum measurements.
\end{abstract}
PACS Codes: 03.65.-w  ; 03.65.Ta  ; 03.65.Ca\\
KEY WORDS:  wave function collapse, caducity of old opinions, Schrodinger's cat, quantum measurements, reconsiderations of quantum theory.
\newpage
\section{Introduction}
A recent highly authorized opinion \cite {1} points out the  existing deadlock that:  \textit{"`There is now ...  no entirely satisfactory interpretation of  Quantum Mechanics "`(QM) }. As  major question of that deadlock is recognized as being  \cite{2}  the problem of Quantum Measurements (QMS), in whose center still stands \cite{3} the Idea about Wave Function Collapse (IWFC). For IWFC, demarcated as above, the most known debates and mainstream publications are reported in \cite{1,2,3}.

Here, in discussing the IWFC question, we try to present a somewhat 'unconventional' strategy based on  viewpoints promoted in our  modest researches about QM,  developed over last few  decades (see \cite {4,5} and references).

Firstly we note the fact that, historically, IWFC emerged at the same time with the inaugural ideas regarding the Conventional Interpretation of Uncertainty Relations (CIUR). In the main CIUR started \cite {4,5} by mixing the theoretical representation (modeling) of a  a physical quantity  regarding a quantum state/system with a \textit{"`fictitious observation"'}(done through some thought (gedanken)  measuring experiment) of the respective quantity. The mentioned mixing  invented and
promoted  the widespread term of \textit{'observable' } for such a  quantity. Below, similarly to the nowadays publications,   we will use also the respective term.

After the alluded start CIUR coagulates in a form of an apparent doctrine centered  on two main pieces : (i) Heisenberg's thought-experimental  formula and (ii) Robertson-Schrodinger theoretical relation. The respective doctrine can be incorporated \cite {4,5} in few basic items  (presumptions/assertions). A deep analysis shows  \cite {4,5} that  the respective items, considered as single or grouped pieces, are incriminated by indubitable  facts which are unsurmountable  within the  framework of CIUR. Then  CIUR proves oneself to be deprived of necessary qualities for a valid scientific construction.  Consequently, in spite of its apology in many modern texts, CIUR must be abandoned as a wrong conception without  any  real value or scientific significance.

In its turn, IWFC  continued to be present in important publications (see \cite{1,2,3} and references), with  explicit or implicit  references  to CIUR. It was aroused by the conflict between two items: (i) the old opinion shortly noted in Abstract   and (ii) the agreement, enforced by theoretical practice, that  studies of quantum systems imply   probabilistic (non-deterministic) entities (wave functions and observables/operators).
For avoiding conflict and breaking a deadlock it was devised the IWFC which, in different readings,  was assumed in a large number of publications. But, as a rule, such assumptions were (and still are) not associated with  adequate investigations regarding the truthfulness of  the respective idea  in relation with the QM  questions. A modest investigation  of that kind we will  try to present below in the next  sections.

Firstly, in Section 2, we point out the fact that in the main (i.e. irrespectively of its readings) IWFC is nothing but an useless fiction. Such a fact certainly shows the caducity and failure of the respective idea. In Section 3 we  discuss the some aspects contiguous between failure of IWFC  and famous subject of Schrodinger's cat thought experiment. Then within Section 4 we argue that alternatively to the IWFC we have to reconsider our views about QM theory in relation with QMS. So, for the  readings of the respective theory, we must to consider either a restricted-QM ($r-QM$)  or an extended-QM ($e-QM$) form. On the one hand the r-QM is essentially the version  promoted by usual QM textbooks \cite{6,7} and it deals exclusively  only with the modeling of intrinsic properties for the studied systems. On the other hand e-QM must to contain also  obligatorily some additional elements regarding QMS descriptions (i.e. theoretical models about characteristics of measuring devices/procedures). Figuratively speaking e-QM consists in r-QM united with QMS descriptions. An simple  exemplification of a QMS description, regarded in the mentioned sense, is presented in the end of the same Section 4. Finally, in Section 5, are given some concluding remarks about the views from this article. 
\section{Uselessness of IWFC }
Now let us try to estimate the usefulness and truthfulness degrees of IWFC. Such an estimation can be obtained if IWFC is  regarded through the details of its constituent elements.  The  before mentioned regard must be opened by  observation  that the starting  purpose of IWFC  was to harmonize the following two conflicting  Items ($\textbf{\textit{I}}$):
\\

$\bullet$ $\textbf{\textit{I}}_1$ : The old opinion  (of the same time as CIUR) that a measurement of a quantum observable $ A $, specific to a state/system at atomic scale,   should be regarded as a single sampling which  gives an unique deterministic  result, say  $a_i$. $\rule{0.15cm}{0.15cm}$
\\

$\bullet$ $\textbf{\textit{I}}_2$: The theoretical agreement that, due to the probabilistic character of wave function $\Psi$ describing the alluded   state/system, the observable $A$ is endowed with a spectrum (set) of distinct values. $\rule{0.15cm}{0.15cm}$
\newpage
So came into an equivocal sight IWFC knew a  lot of debates (see \cite{1,2,3} and references). In essence, the solution promoted by the respective debates can be summarized within the following Subterfuge ($\textbf{\textit{S}}$):
\\

$\bullet$ $\textbf{\textit{S}}$ : The  unique  result $a_i$ and wave function $\Psi$,   mentioned in items $\textbf{\textit{I}}_1$ and  $\textbf{\textit{I}}_2$,  should be seen ( and described) through the wave function  collapse $\Psi\longmapsto\psi_i$, where $\Psi$ depicts the considered  quantum state/system in its wholeness while $\psi_i$ is  the $a_i$-eigenfunction of the operator $\widehat{A}$ (associated to the observable  $ A $) - i.e $\widehat{A}\psi_i=a_i\psi_i$. $\rule{0.15cm}{0.15cm}$
\\
For a proper judgment of such a subterfuge we have to reconsider  the correctness of the items $\textbf{\textit{I}}_1$ and $\textbf{\textit{I}}_2$. In the light of such a reason it must to  note  that  studies  from the last decades (see\cite{4,5,6,7} and references)  consolidated beyond doubt the fact that, mathematically, a quantum observable $A$ (through of the operator $\widehat{A}$ ) is a true random variable. In a theoretical viewpoint, for a given quantum state/system,   such a variable is regarded as endowed with a  spectra of values associated with corresponding  probabilities (more exactly probability amplitudes). Then, from an experimental perspective, a measurement of a quantum observable requires an adequate number  of samplings  finished through a  significant statistical group of data (outcomes).

Previous  opinions about the randomness  of quantum observables can be consolidated indirectly by  mentioning the  quantum-classical probabilistic similarity (see \cite{4,8})  among the respective observables and macroscopic variables  studied within  phenomenolgical (thermodynamic) theory of fluctuations \cite{4,9,10,11,12,13,14}. In this way let us refer to such a macroscopic random observable $\widehat{\mathbb{A}}$.  Its  intrinsic ($in$) characteristics are given in details by  a  continuous  spectra of values $\mathcal{A}$ inside  of    spectra  (range) $\Omega_{in}$  (i.e. $\mathcal{A}\in\Omega_{in}$ ),  associated  with a probability density  $w_{in} = w_{in}(\mathcal{A})$ . Then for $\widehat{\mathbb{A}}$, in its fullness, a single experimental sampling delivering an unique (individual) result, say $\mathcal{A}_i$,  is worthlessly.  Such a sampling is not described as a collapse of the probability density $w_{in}(\mathcal{A})$. Moreover a true  experimental evaluation of $\widehat{\mathbb{A}}$, in its wholeness and regarded  equivalently with  a stationary random process,  requires \cite{15} an adequate  lot of samplings finished through a  significant statistical set of individual  recordings. In a plausible modeling \cite{16,17} the mentioned recordings ($rec$) can be described by another probability density $w_{rec} = w_{rec}(\mathcal{A})$. 

 The above notifications about quantum observables point out clearly the complete incorrectness of item $\textbf{\textit{I}}_1$. Consequently, even if in the main the item $\textbf{\textit{I}}_2$ is a true assertion, the subterfuge $\textbf{\textit{S}}$ supporting IWFC proves oneself to be nothing but an useless recommendation. Additionally note that, in the mainstream of publications ( see \cite{1,2,3} and references), the respective subterfuge is not fortified with  thorough (and  genuine)   descriptions regarding  the collapse $\Psi\longmapsto\psi_i$.  Evidently that the above revealed facts \textbf{\textit{ point out the caducity and failure  of IWFC}}.
\\

The previous discussions about IWFC lead us also to the following more general Remark ($ \textit{\textbf{R}}$)
\\

$\bullet$ $\textit{\textbf{R}}$:  A random variable  should not be assessed ( measured ) by an unique deterministic sampling (trial) but   by a statistical  ensemble of samplings.$\rule{0.15cm}{0.15cm}$

\section{Contiguities with the  Schrodinger's cat\\
 thought experiment}
As it is well known \cite{18} the famous  Schrodinger's cat thought experiment is a subject often displayed in debates (more or less scientifically) about the significance/interpretations of QM constituents.   The essential element in the respective experiment  is represented by a killing single decay of a radioactive atom. But the radioactive decays are random (probabilistic) events. Then the mentioned killing decay is in fact a twin analogue of the single sampling noted above  in item $\textbf{\textit{I}}_1$ in connection with IWFC.

The mentioned analogy motivates us to discuss on some contiguities among  questions specific to the   alluded experiment and those regarding IWFC.  We think that, according to the above remark $\textit{\textbf{R}}$,   the main point of such motivated  discussions is to mark down the following Notification ($ \textit{\textbf{N}}$)
 \\
   
$\bullet$ $ \textit{\textbf{N}}$ : When the variable of interest has random characteristics it is useless (even forbidden) to design experiences or actions that relies solely on a single deterministic sampling of that variable.$\rule{0.15cm}{0.15cm}$ 
\\

In the light of such notification the Schrodinger's experiment appears to be  noting  but just a  
fiction ( figment ) without any scientific  value. That is why the statements like :\textit{"`the Schrodinger cat thought experiment remains a topical touchstone for all interpretations of  quantum mechanics"'}, must be regarded as being worthlessly. (Note that such statements are present in many science popularization texts, e.g. in the ones disseminated via \textit{INTERNET})

The above notification $ \textit{\textbf{N}}$, argued for quantum level,  can be   also   of non-trivial significance (interest)   at macroscopic scale. For illustrating such a significance let us refer to the thought  experimental situation of a classical  (macroscopic) cousin of the Schrodinger's cat. The regarded situation can be depicted as follows. The cousin is placed in a sealed box together a flask of poison and an internal macroscopic actuator. The actuator is  connected to an macroscopic uncontrollable (unobservable) sensor located within the circular error probable (CEP) of a  ballistic  projectile trajectory. Note that a ballistic projectile is a missile only guided during the relatively brief initial powered phase of flight, whose course is subsequently governed by the laws of classical mechanics. CEP is defined as the radius of a circle, centered about the mean, whose boundary is expected to include the landing points of 50\verb'%' of the launching rounds (for more  details about ballistic terminology see \cite{19}). The experiment consists in launching of  a single  projectile,   without any possibility to observe the point  where it  hits the ground. Additionally the projectile is equipped with a  radio transmitter  which signals the moment of impact and therefore the flight time.  If the sensor is smitten   by  projectile   the actuator is activated releasing the poison that kills the cousin. But as the  projectile trajectory has a probabilistic character (mainly due to the external ballistic factors) the hitting point is placed with the probability of 50\verb'%'  within the surface of CEP where the sensor is located. That is why, after the projectile time of flight and  without opening the box, one  can not know the  state of living for the cousin. So the whole situation of the  classical cousin is completely analogous with the one of quantum Schrodinger's cat. Therefore the thought experiment with classical cousin makes evident oneself as another fiction without any
real significance.

Besides the mentioned example with Schrodinger's cat cousin  we can add here another circumstance where the above notification $ \textit{\textbf{N}}$ is taken into account (and put in practice) in a classical context. Namely we think that, in the last analysis, the respective notification  is the deep reason of the fact that in practice of the traditional artillery (operating only with ballistic projectiles but not with propelled missiles)   for  destroying a military objective  one uses a considerable (statistical) number of projectiles but not a single one.
\newpage  
\section{Contiguities with descriptions of\\ quantum measurements}
It is easy to see the fact that the  considerations from Section 2 are contiguous with the question of QMS descriptions. Such a fact require directly   certain additional comments which we try to present  here below. In our opinion the mentioned question must be regarded  within a context marked by the following set of Topics (\textbf{\textit{T}}):
\\

$\bullet$ $\textbf{\textit{T}}_1$ : In its plenitude the  QM theory must be considered in a r-QM respectively in an e-QM reading. Fundamentally r-QM deals with theoretical  models regarding intrinsic properties of quantum (atomically sized) systems. On the other hand e-QM has  to take into account both the characteristics of measured observable/system and the  peculiarities of measuring devices/procedures. $\rule{0.15cm}{0.15cm}$
 \\
 
 $\bullet$  $\textbf{\textit{T}}_2 $: Within r-QM a state of a system is described completely by its intrinsic ($in$) wave function $\Psi_{in}$ and operators $\widehat{A_k}$ ($k = 1,2, ...,f$), associated to its specific  observables $A_k$. Note that  expression of $\Psi_{in}$ is distinct for each situation (state/system) while the operators $\widehat{A_k}$ have the same mathematical representation  in many situations. In addition
 the concrete mathematical  expression for $\Psi_{in}$ may be obtained either from theoretical  investigations (e.g. by solving the adequate Schrodinger equation) or from  a priori  considerations (not supported by factual studies).  For a given state/system  the observables  $A_k$ can be put into sight through  a small number of global $in$-descriptors such are: $in$-mean/expected values, $in$-deviations or second/higher order $in$-moments/correlations (for few examples see below). $\rule{0.15cm}{0.15cm}$
\\

 $\bullet$ $\textbf{\textit{T}}_3$: A true experimental evaluation of  quantum observables can be obtained by means of an adequate  numbers of  samplings finished through  significant statistical sets of individual  recordings. For an observable the samplings must be done on the same occurrences ( i.e. practically on  very images of the investigated observable and state/system). As regards a lot of observables a global and easy sight of the mentioned evaluation can be done by computing from the alluded recordings some (experimental-) $exp$-quantifiers (of global significance) such are : $exp$-mean , $exp$-deviation respectively $exp$- higher order moments.  $\rule{0.15cm}{0.15cm}$
\newpage

$\bullet$ $\textbf{\textit{T}}_4$:  Usually,  a first confrontation of theory   versus experience, is  done by comparing side by side the $in$-descriptors and $exp$-quantifiers mentioned  above  in $\textbf{\textit{T}}_2$ and $\textbf{\textit{T}}_3$. Then, if the  confrontation is confirmatory, the investigations about the studied observable/system can be  noticed as a fulfilled task. If the alluded confirmation does not appear the investigations may be continued by resorting to one or groups of the following  upgradings ($\textbf{\textit{u}}$) :\\
$\textbf{\textit{u}}_1$ : An amendment for expression of  $\Psi_{in}$ , e.g. through solving a more complete Schrodinger equation or using the  quantum perturbation theory.\\
$\textbf{\textit{u}}_2$ : Improvements of experimental devices/procedures.\\
$\textbf{\textit{u}}_3$ : Addition of a  theoretical description for the considered  QMS.  $\rule{0.15cm}{0.15cm}$
\\

 $\bullet$  $\textbf{\textit{T}}_5 $: Through the extension suggested in above upgrading $\textbf{\textit{u}}_3$ the study changes its reading from a r-QM into an e-QM vision, in the sense mentioned in topic $\textbf{\textit{T}}_1 $. Such an extension needs to be  conceived as a stylized representation through a  schematic/mathematic  modeling so that  it to  include both intrinsic  elements  (regarding observables/states/systems) and measuring details (about devices/procedures). Additionally if the upgrading $\textbf{\textit{u}}_3$ is adopted  then a true confrontation of theory   versus experience must be done not as it was mentioned in $\textbf{\textit{T}}_4 $  but by   putting face to face the predictions of QMS description with the experimental data $\rule{0.15cm}{0.15cm}$
\\

For an  illustration of the  topics $\textbf{\textit{T}}_1$ - $\textbf{\textit{T}}_5$
let us  regard as a QM system a spin-less quantum  particle in a rectilinear  and stationary movement along the $Ox$ axis . The QMS problems   will be reported to  the  orbital  observables momentum $p_x$ and energy $E$, denoted generically by  $A$.

In terms of $\textbf{\textit{T}}_2$  the probabilistic intrinsic ($in$)   characteristics of such particle  are depicted by orbital wave function $\Psi_{in}=\Psi_{in}(x) $ (where coordinate $x$  covers  the range $\Omega$). The observables $A$ are described by the associated operators $\widehat{A}$ according the QM rules \cite{6,7} (i.e. by $\widehat p_x {\kern 1pt}  = {\kern 1pt} \, - i\hbar \frac{\partial }{{\partial x}}$ respectively by the Hamiltonian $\widehat{H}$  ). Then from the class of global $in$-descriptors regarding  such an observable $A$ can be mentioned the $in$-mean-value
 $ \left\langle A \right\rangle _{in} $ and $in$- deviation  $\sigma _{in} \left( A \right)$ defined as follows
\begin{equation}\label{eq:1}
\left\langle A \right\rangle _{in}  = \left( {\Psi _{in} ,\;\widehat A\;\Psi _{in} } \right)\quad ;\quad \sigma _{in} \left( A \right) = \sqrt {\left( {\delta _{in} \widehat A\;\Psi _{in} ,\;\delta _{in} \widehat A\;\Psi _{in} } \right)} 
\end{equation} 
where $(f,g)$ denotes the scalar product of functions $f$ and  $g$, \\
while $\delta _{in} \widehat A = \widehat A - \left\langle A \right\rangle _{in} $

An  actual experimental measurement of observable $A$ in sense of $\textbf{\textit{T}}_3$ must be done through a set of statistical samplings. The   mentioned set gives for $A$ as recordings a collection of distinct values $ \left\{ {\alpha _1 {\kern 1pt} ,\alpha _2 ,\alpha _3 {\kern 1pt} ,\,...\,,\alpha _r {\kern 1pt} {\kern 1pt} } \right\} $ associated with the  empirical probabilities (or relative frequencies) 
$\left\{ {\nu _1 {\kern 1pt} ,\nu _2 ,\nu _3 {\kern 1pt} ,\,...\,,\nu _r {\kern 1pt} {\kern 1pt} } \right\}$.
Usually, for a  lower  synthesized sight about the mentioned measurement, as   experimental ($exp$) quantifiers  are chosen the  $exp$-mean $\left\langle A \right\rangle _{\exp } $ and 
$exp$-deviation $\sigma _{\exp } \left( A \right)$ given through the formulas:
\begin{equation}\label{eq:2}
\left\langle A \right\rangle _{\exp }  = \sum\limits_{j = 1}^r {\nu _j  \cdot \alpha _j } \quad ;\quad \sigma _{\exp } \left( A \right) = \sqrt {\sum\limits_{j = 1}^r {\nu _j  \cdot \left( {\alpha _j  - \left\langle A \right\rangle _{\exp } } \right)^2 } } 
\end{equation}

The above considerations about an experimental QMS must be supplemented with the following Observations (\textit{\textbf{O}}) :
\\
\\
$\ast$  $ \textit{\textbf{O}}_1$ : Note that due to the inaccuracies of experimental devices
 some of the recorded values  $ \left\{ {\alpha _1 {\kern 1pt} ,\alpha _2 ,\alpha _3 {\kern 1pt} ,\,...\,,\alpha _r {\kern 1pt} {\kern 1pt} } \right\} $ can differ from the eigenvalues
 $\left\{ {a_1 {\kern 1pt} ,a_2 ,a_3 {\kern 1pt} ,\,...\,,a_s {\kern 1pt} {\kern 1pt} } \right\}$ of the operator $\widehat{A}$.$\blacktriangle$ \\ 
$\ast$  $\textbf{\textit{O}}_2$ : A comparison  at first sight between  theory and experiment can be done by putting   side by side the corresponding aggregate (global) entities  \eqref{eq:1} and \eqref{eq:2}. When  one finds that the values of compared entities are in   near equalities,  usually is admitted  the following couple of linked  beliefs ($\textbf{\textit{b}}$) :\\
 ($\textbf{\textit{b}}_1$) theory is pretty correct and \\
 ($\textbf{\textit{b}}_2$) measuring devices/procedures are almost ideal. \\
 Thus, practically, the  survey of debated QMS can be regarded as a finished task. $\blacktriangle$\\
$\ast$ $\textit{\textbf{O}}_3$ : If instead of the mentioned equalities one detects (one or two) flagrant differences at least one of the alluded beliefs ($\textbf{\textit{b}}_1$) and ($\textbf{\textit{b}}_2$)   is deficient  (and unsustainable). Such a deadlock can be avoided by one or groups  of the upgradings $\textbf{\textit{u}}_1$ - $\textbf{\textit{u}}_3$ mentioned above within the topic $\textbf{\textit{T}}_4$. $\blacktriangle$
\\

Generally speaking the the upgradings $\textbf{\textit{u}}_1$ - $\textbf{\textit{u}}_2$ are appreciated and worked (explicitly or implicitly) in mainstream literature (see \cite{1,2,3} and references). But note that, as far as know, for $\textbf{\textit{u}}_3$ such an appreciation was neither taken into account  nor developed in details in the respective literature. It is our modest task  to present below  a brief  exemplification of upgrading $\textbf{\textit{u}}_3$ in relationship with the QMS question.    The presentation is done in some simple terms of  information transmission theory. 
 
 \subsection*{An information theory modeling for QMS description} 
In  a QMS process the input  information regarding the intrinsic ($in$) properties of the measured system is converted in predicted ($pd$) or output information incorporated within the data received on a device recorder. That is why a QMS appears as an \textit{ information  transmission process } in which the measuring device plays the role of a \textit{ information transmission channel}.  So  the QMS   considered above can be symbolized as $\Psi _{in}  \Rightarrow \Psi _{pd} $ for the wave function  while the operator $\widehat{A}$ remains invariant. Such symbolization is motivated by the facts that, on the one hand the wave function $\Psi$ is specific for each considered situation (state/system) whereas, on the other hand the operator $\widehat{A}$ preserves the same mathematical expression in all (or at least in many) situations. Note that the (quantity of) information is connected with probability densities $\rho_{\eta}(x)$ and currents (fluxes) $j_{\eta}(x)$ ($\eta= in,pd$) defined in terms of $\Psi_{\eta}(x)$ as in usual QM \cite{4,5,6,7}. Add here the fact that $\rho _\eta  \left( x \right)$  and $j_\eta  \left( x \right)$ refer to the positional respectively  the motional kinds of probabilities . Experimentally the two kinds of probabilities can be regarded as measurable by distinct devices and procedures. Besides, as in practice, one can suppose that the alluded devices are stationary and linear. Then, similarly with the case of measurements regarding classical random observables  \cite{4,16,17}, in an informational reading , the  essence of here discussed QMS description  can be compressed  \cite{4,17} through the relations:
\begin{equation}\label{eq:3}
\rho _{pd} \left( x \right) = \int {\Gamma \left( {x,x'} \right)} \,\rho _{in} \left( {x'} \right)dx'\;;\quad j_{pd} \left( x \right) = \int {\Lambda \left( {x,x'} \right)} \,j_{in} \left( x \right)dx'
\end{equation}
Here the kernels $\Gamma(x,x')$ and $\Lambda(x,x')$ include as noticeable parts some elements about  the peculiarities of measuring devices/procedures. Mathematically,  $\Gamma(x,x')$ and $\Lambda(x,x')$ are normalized in respect with both $x$ and $x'$. Note that QMS becomes nearly ideal when   both $\Gamma(x,x')\rightarrow\delta(x-x')$ and $\Lambda(x,x')\rightarrow\delta(x-x')$,  ( $\delta(x-x')$ being  the Dirac's $\delta$ function). In all other cases QMS appear as non-ideal. 

By means of the probability density $\rho_{pd}(x)$ and current $j_{pd}(x)$  can be computed \cite{4}  some useful expressions like $\Psi _{pd}^* \left( x \right)\widehat A\,\Psi _{pd} \left( x \right)$. Then, for observable $A$, it is possible to evaluate global indicators  of predicted ($pd$) nature such are $pd$-mean 
$\left\langle A \right\rangle _{pd}$ and $pd$-deviation $\sigma _{pd} \left( A \right)$  defined,  similarly with \eqref{eq:1},  as follows
\begin{equation}\label{eq:4}
\left\langle A \right\rangle _{pd}  = \left( {\Psi _{pd} ,\widehat A\Psi _{pd} } \right)\quad ;\quad \sigma _{pd} \left( A \right) = \sqrt {\left( {\delta _{pd} \widehat A\Psi _{pd} ,\delta _{pd} \widehat A\Psi _{pd} } \right)} 
\end{equation} 

If as regards  a quantum observable $A$, besides a true experimental evaluation, for its measuring process one resorts to a (theoretical/informational) QMS description of the above kind the $pd$-indicators \eqref{eq:4} must be tested by comparing them with their experimental (factual) correspondents (i.e. $exp$- quantifiers) given in \eqref{eq:2}. When the test is confirmatory both theoretical descriptions, of r-QM intrinsic properties of system/state  respectively of QMS, can be considered as adequate and therefore the scientific task can be accepted as finished. But, if the alluded test  is of invalidating type,  at
least one of the mentioned descriptions must be regarded as inadequate and the whole question requires further investigations.

For an impressive  illustration of the above presented informational QMS description  we consider as observable of interest the  energy $A =E= H$ regarding a QM harmonic oscillator. The operator $\widehat H$ associated to the respective observable is 
$\widehat H =  - \frac{{\hbar ^2 }}{{2m}}\frac{{d^2 }}{{dx^2 }} + \frac{1}{2}m\,\omega ^2 x^2 $ ($m$ and $\omega$ denote the mass respectively the angular frequency of oscillator). The  oscillator is considered to be in its lower energetic level, whose intrinsic state is described by the wave function $\Psi _{in} \left( x \right) \propto \exp \left\{ { - \frac{{x^2 }}{{4\sigma ^2 }}} \right\}$ (here $\sigma  = \sigma _{in} \left( x \right) = \sqrt {\frac{\hbar }{{2m\omega }}} $ denote the $in$-deviation of coordinate $x$ ). Then, because    $\Psi_{in}$ is a real function, for the considered state one finds  $j_{in}=0$ -  i.e. the probability current is absent. So for the regarded QMS description in \eqref{eq:3} remains of interest only first relation dealing with the change $\rho_{in}\rightarrow\rho_{pd}$ of the probability density through the kernel $\Gamma(x,x')$. If the supposed measuring device has high performances  $\Gamma(x,x')$  can be taken \cite{4} of Gaussian form 
 i.e. $\Gamma \left( {x,x'} \right) \propto \exp \left\{ { - \frac{{\left( {x - x'} \right)^2 }}{{2\gamma ^2 }}} \right\}$,  $\gamma$  being the  error characteristic of the respective device. It can been seen that in the case when  $\gamma\rightarrow 0$ the kernel $\Gamma(x,x')$ degenerates into the Dirac function $\delta(x-x')$. Then $\rho_{pd}=\rho_{in}$. Such a case corresponds to an ideal measurement. Differently,  the cases when $\gamma\neq0$  are associated with non-ideal measurements.

In the above  modeling of QMS description for the energy $A =E= H$ one obtains \cite{4} the following $in$- respectively $pd$- means and deviations
\begin{equation}\label{eq:5}
\left\langle H \right\rangle _{in}  = \frac{{\hbar \omega }}{2}\quad ;\quad \sigma _{in} \left( H \right) = 0
\end{equation}
\begin{equation}\label{eq:6}
\left\langle H \right\rangle _{pd}  = \frac{{\omega \left[ {\hbar ^2  + \left( {\hbar  + 2m\omega \gamma ^2 } \right)^2 } \right]}}{{4\left( {\hbar  + 2m\omega \gamma ^2 } \right)}}
\end{equation}
\begin{equation}\label{eq:7}
\sigma _{pd} \left( H \right) = \frac{{\sqrt 2 m\omega ^2 \gamma ^2 \left( {\hbar  + m\omega \gamma ^2 } \right)}}{{\left( {\hbar  + 2m\omega \gamma ^2 } \right)}}
\end{equation}
Relations \eqref{eq:5} and \eqref{eq:7} show that even if  $\Psi_{in}$ has the quality of an eigenfunction for $\widehat H$  (as $\sigma_{in} (H) =0$), due to the measurement  $\Psi_{pd}$ is deprived of such a quality (because $\sigma_{pd}(H) \neq0$).

 \section{Concluding remarks }
We point out, on the one hand the historical emergence of the IWFC from the conflict between the items $\textbf{\textit{I}}_1$  and  $\textbf{\textit{I}}_2$  mentioned in Section 2. Then  we remind the fact that, on the other hand, the modern studies certify the random characteristics of quantum observables. Therefore  a true  measurement of such an observable requires a whole set of statistically  significant  samplings. The respective requirement invalidate  indubitably the alluded item $\textbf{\textit{I}}_1$ . So IWFC is proved as a caducous and useless recommendation.  

Contiguously the respective proof allows to put into a new light the famous Schrodinger's cat thought experiment. We argue in Section 3 that Schrodinger's  experiment is noting a but just a fiction  without any scientific value. The argumentation relies on the notification that:"' When the variable of interest has random characteristics it is useless (even forbidden) to design experiences or actions that relies solely on a single deterministic sampling of that variable"'. The same notification is useful in appreciating of some non-quantum problems  such are  a Schrodinger's-type experiment with a classical cat or statistical practices in traditional artillery .

The question of IWFC  caducity  is contiguous also with the problem of QMS descriptions. That is why in Section 4  we present some brief considerations about  the respective problem. Thus we propose that QM theory to be regarded  either in a r-QM or in an e-QM reading, as it refers to the studied observables/systems without or with taking into account the QMS descriptions. The proposal is consolidated
 with  simple  illustration. Particularly we suggest an  approach of QMS descriptions based on information transmission theory .
 
Of course that other  different approaches about QMS descriptions can be imagined. They can be taken into account  for extending QM theory towards an e-QM reading, as complete/convincing as possible.

\end{document}